\documentclass[usenatbib]{basi}
\usepackage[british]{babel}
\usepackage[varg]{txfonts}
\usepackage{rotating}
\usepackage{dcolumn}

\def \mnras {MNRAS}
\def \apj {ApJ}
\def \apjs {ApJS}

\def \aap {A\&A}

\def \angstrom {{\rm \AA}}

\begin{document}
\title[Abundance anticorrelations on integrated spectra]
{Modelling chemical abundance anticorrelations on globular cluster spectra}
\author[P.~Coelho et~al.]%
       {P.~Coelho$^1$\thanks{email: \texttt{paula.coelho@cruzeirodosul.edu.br}},
       S.~Percival$^{2}$ and M.~Salaris$^2$\\
       $^1$N\'ucleo de Astrof\'{\i}sica Te\'orica, Universidade Cruzeiro do Sul, 
R. Galv\~ao Bueno 868, 01506-000, S\~ao Paulo, Brasil\\
       $^2$Astrophysics Research Institute, Liverpool John Moores
University, 12 Quays House, Birkenhead, CH41 1LD, UK}

\pubyear{2011}
\volume{00}
\pagerange{\pageref{firstpage}--\pageref{lastpage}}

\date{Received \today}

\maketitle
\label{firstpage}

\begin{abstract}
It is widely accepted that individual Galactic globular clusters harbor two coeval 
generations of stars, the first one born with the `standard' $\alpha$-enhanced metal mixture 
observed in field Halo objects, the second one characterized by an anticorrelated CN-ONa abundance 
pattern overimposed on the first generation, $\alpha$-enhanced metal mixture. 
We have investigated with appropriate stellar population synthesis models
how this second generation of stars affects the integrated spectrum of a typical metal rich 
Galactic globular cluster, like 47\,Tuc. 
Our main conclusions are: 1) the age-sensitive Balmer line,  
Fe line and the [MgFe] indices widely used to determine age, 
Fe and total metallicity of extragalactic systems are largely insensitive to the 
second generation population; 
2) enhanced He in second generation stars 
affects the Balmer line indices of the integrated spectra,  
through the change of the turn off temperature and the 
horizontal branch morphology of the underlying isochrones, which translate into a bias 
towards slightly younger ages. 
\end{abstract}


\section{Introduction}\label{s:intro}

A large body of spectroscopic data published in the last 10 years has conclusively established 
the existence of primordial surface chemical abundance variations of C, N, O, Na  
in individual Galactic 
globular clusters (GCs) of all metallicities, from 
very metal rich objects like NGC6441, to the most metal poor ones like M15 \citep[see, e.g.,][]{cbg10}. 

The accepted working scenario to explain these CNONa anticorrelations 
prescribes that the stars currently evolving in a GC were born with the observed CNONa patterns. 
Intermediate-mass asymptotic giant branch (AGB) stars in the range $\sim 3-8 M_{\odot}$ 
(and/or the slightly more massive 
super-AGB stars) and massive rotating stars are considered viable sources of the necessary heavy-element 
pollution \citep[see, e.g.,][]{vd05}. 
Slow winds from the envelopes of AGB (or super-AGB stars) or from the equatorial disk formed around fast rotating 
massive stars inject matter into the intra-GC medium after a time of order $10^6 - 10^8$~yr, depending on 
the polluter. Provided that a significant fraction of the material is not lost 
from the cluster, new stars (second generation, but essentially coeval with the 
polluters' progenitors, given their short evolutionary timescales) may be able to form directly out of pristine gas  
polluted to varying degrees by these ejecta, that will show the observed anticorrelation patterns.
Another by product of this pollution may possibly be 
an enhanced initial He-abundance for these second generation stars. 
Recent numerical models by \citet{derc10} have started to explore in quantitative details this broad picture.


\section{Models}

We have started by modelling a reference set of $\alpha$-enhanced SSPs  -- metal mixture with 
[$\alpha$/Fe]$\sim$ 0.4 -- with ages t=12 and 14 Gyr, which we identify hereafter as \emph{standard models}.
We have considered in this
investigation a single [Fe/H]=$-$0.7 ($Z$=0.008), 
typical of the metal rich subpopulations of Milky Way -- and of 47Tuc, whose integrated spectrum 
is a benchmark for population synthesis models -- and an initial He mass fraction $Y$=0.256.
For both ages we have then considered a second generation population (\emph{modified CNONa models}) whose metal composition has  
C decreased by 0.30~dex, N increased by 1.20~dex, O decreased by 0.45~dex and Na increased by 0.60~dex with 
respect to the first generation $\alpha$-enhanced mixture, all other metal abundances  
being unchanged. This pattern is typical of values close to the upper end of the observed anticorrelation 
patterns in Galactic GCs \citep{c09}. 
The metal distribution of this second generation coeval population has the same C+N+O sum 
and the same Fe abundance (as a consequence also the total metallicity $Z$ will be practically the same) 
as the first generation composition, in agreement with spectroscopic measurements on 
second generation stars within individual Galactic GC.
For both the 12 and 14~Gyr second generation populations we have accounted for two alternative 
values of $Y$, e.g. $Y$=0.256 -- as in the first stellar generation -- and $Y$=0.300, to include a 
possible enhancement of He in second generation stars. This latter, hereafter identified as \emph{He enhanced models}, is consistent with constraints on the typical enhancement of He in Galactic GCs, 
as determined by \citet{bragag10}. 

We stress that these choices 
for the chemical composition of second generation populations are very general, and 
should give us a realistic estimate of their impact on GC analyses using SSPs. 
Overall, our selection of representative extreme values of the CNONa variations 
will provide a first important indication of the maximum effect
of these abundance anomalies on GC integrated spectra, and will serve as a guideline to interpret 
the abundance pattern derived from fitting SSPs to the observed spectra of Galactic and extragalactic GCs.

The SSP models were computed adopting isochrones from BaSTI \citep{basti06} and a stellar synthetic grid computed with
ATLAS12 \citep{kur81,cas05} and SYNTHE \citep{kur81,sbo04} codes. One synthetic grid was computed for each set of models -- reference $\alpha$-enhanced, second gereration with Y=0.256 and second generation with Y=0.300 -- properly taking the respective chemical mixture into account. From the appropriate isochrone and grid of synthetic stellar spectra  
we have calculated the integrated spectrum for these first and second generation populations, 
employing the \citet{kr01} initial mass function. The integrated spectra modelled cover the 
wavelength region 3500 to 6000\,${\rm \AA}$ at a spectral resolution $R=\lambda/\delta\lambda=10000$.
We refer the reader to \citet{coelho+11a} for more details on our models.

We show in Fig. 1 the spectral models for 12\,Gyr populations. 

\begin{figure}
\centerline{\includegraphics[bb=55 430 515 710,width=11cm]{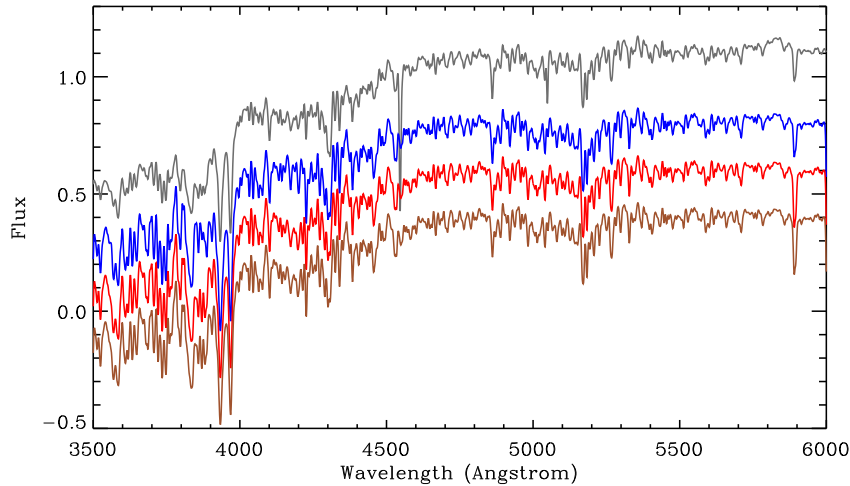}}
\caption{The spectral stellar population models computed for this investigation are shown in blue, red and brown lines (standard, modified CNONa and He-enhanced modes, respectively -- see text for details). Models shown are those for 12\,Gyr. In gray line in shown the observed integrated spectrum of 47Tuc from \citet{schiavon+05}. The model spectra were convolved to match the spectral resolution of the observation, FWHM $\sim$\,3.1\,\angstrom.
}
\end{figure}


\section{Results}

In Figs. 2 to 4 we zoom some regions of the spectra -- known 
to be sensitive to C, N, Na and atomic O -- to highlight the effect (or absence of an effect) of the abundance change on 
the stellar population spectra. In the case of C and N (Fig. 2) and Na (Fig. 4), the effect is clearly seen in the spectra.
In the case of atomic oxygen, the effect is so small (lower than 0.5\% in flux) that the two models are indistinguishable. 

\begin{figure}
\centerline{\includegraphics[bb=55 505 515 710,width=11cm]{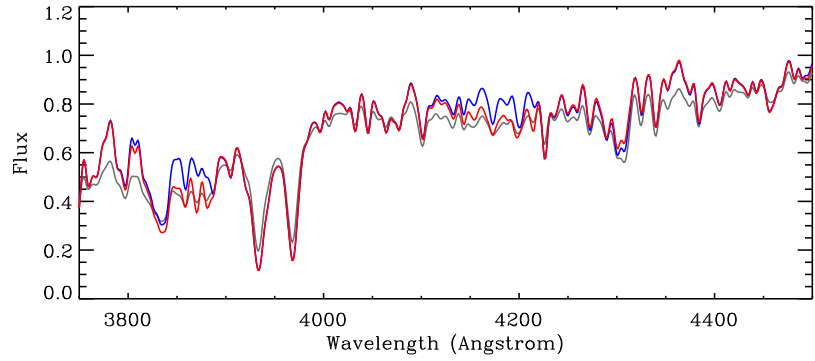}}
\caption{Standard and modified CNONa models are shown in blue and red respectively, for a spectral region sensitive to molecular carbon and nitrogen. Observed 47Tuc spectrum is shown in dark gray.}
\end{figure}

\begin{figure}
\centerline{\includegraphics[width=11cm]{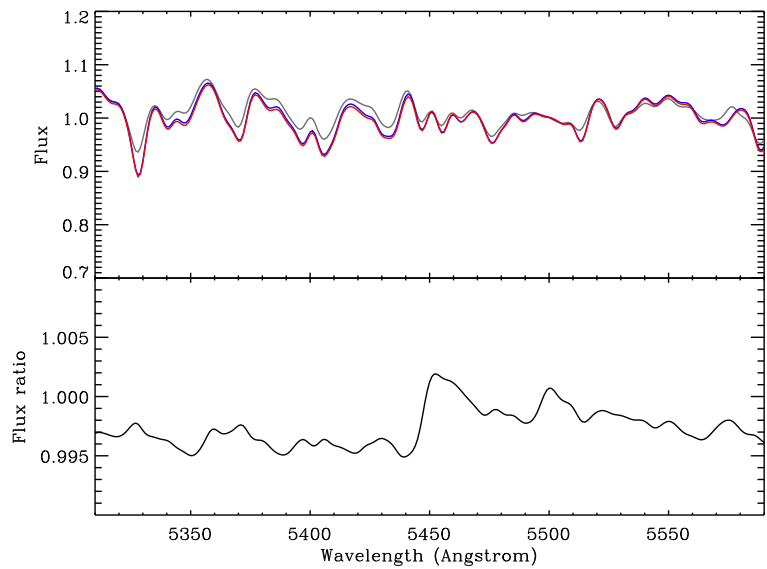}}
\caption{Same as in Fig. 2, for a spectral region sensitive to atomic oxygen. The ratio between the standard and modified CNONa model is shown in the bottom panel.}
\end{figure}

\begin{figure}
\centerline{\includegraphics[bb=55 360 530 550,width=11cm]{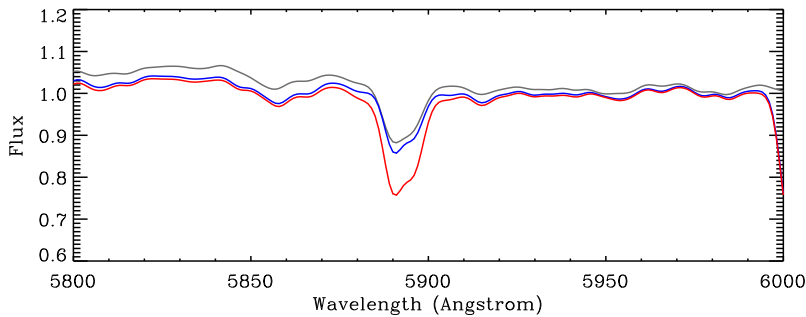}}
\caption{Same as in Fig. 2, for a spectral region sensitive to sodium.}
\end{figure}

The effect of the CNONa modified abundances on Lick/IDS spectral indices
has been presented in detail in \citet{coelho+11a}.
To summarize, the presence of a second generation of GC stars with unchanged He-content affects appreciably 
only the Ca4227, G4300, ${\rm CN_1}$, ${\rm CN_2}$ and NaD metal indices. 
Very importantly, the variation of Ca4227 goes in the direction to mimic a lower 
Ca abundance, if this index is used as a measure of the Ca content. 
Our results support \citet{lw05} suggestion that 
the effect of a second generation population with CNONa anticorrelations 
may explain the discrepancies they find when comparing 
Ca4227, ${\rm CN_1}$, ${\rm CN_2}$ and NaD index strength (on the Lick IDS system) from their 
$\alpha$-enhanced SSP models, with data for Galactic and M31 GCs.
On the other hand, age, Fe-abundance and $Z$  
inferred from ${\rm H\beta}$-Fe5406 and ${\rm H\beta}$-[MgFe] diagrams (or the ${\rm H\delta_F}$ and,  
to a slightly smaller extent, the ${\rm H\gamma_F}$ counterpart) are confirmed once again to be robust 
and insensitive to the chemical abundance pattern of second generation stars. 

When we consider the case of a second generation population with enhanced 
He-abundance ($Y$=0.300), the values of all metal indices in this second generation population 
are essentially identical to the $Y$=0.256 case, but the Balmer line indices are changed by the increase of He, mimicking an age younger by 2\,Gyr. We detected that the changes in ${\rm H\beta}$, ${\rm H\delta_F}$ 
and ${\rm H\gamma_F}$ compared to the $Y$=0.256 case are due to differences in the underlying isochrone  
representative of the He-enhanced second generation population. 
The TO and main sequence of the underlying isochrone are hotter by $\sim$100~K 
compared to the $Y$=0.256 isochrone. Given that the TO mass at fixed age is 
smaller (by about 0.07~${\rm M_{\odot}}$ in our case) in the 
He-enhanced isochrone, and we assume that the total mass lost along the RGB is the same (0.11~${\rm M_{\odot}}$), the 
mass evolving along the HB is smaller, hence the typical ${\rm T_{eff}}$ of HB stars is higher, again by $\sim$100~K. 
These higher ${\rm T_{eff}}$ values for TO and HB increase the value of the Balmer line indices.

It remains yet to be studied what is the effect of modified CNONa (and He) abundances on the metallicites and ages
derived via spectral fitting codes \citep[e.g][]{ulyss}. On the other hand, one might test the possibility to measure
abundances -- beyond iron and $\alpha$-elements -- with spectral fitting (e.g. by using these models to
differentially correct stellar populations, see \citealt[][]{walcher+09}). 

It is likely possible to detect the existence of those abundance
changes in the residuals of a spectral fitting analysis. At a first approximation, we could assume the residuals due a significant change 
on the CNONa abundance will show a pattern similar to the differential spectrum between our standard and modified CNONa models, shown in Fig. 5.  The signal of C and N are clearly seen, as the modified CNONa abundances enhances the CN features by up to 30\% in flux. In fact, primordial abundance variations such as the ones explored in this modelling might be an explanation for the long term problem of the particularly strong CN bands in globular clusters in M31, when compared to the ones of the Milky Way \citep[e.g.][]{burstein+84,trager04}. In Fig. 6 is shown the ratio between the He enhanced and modified CNONa models, to illustrate the effect of enhancing He alone. We note that features related to the hydrogen lines and the small change of the continuum slope, characteristic of an integrated "hotter" model, are in agreement with the explanation above on the effect of the He enhancement on the TO and HB temperatures.

\begin{figure}
\centerline{\includegraphics[bb=55 360 530 550,width=11cm]{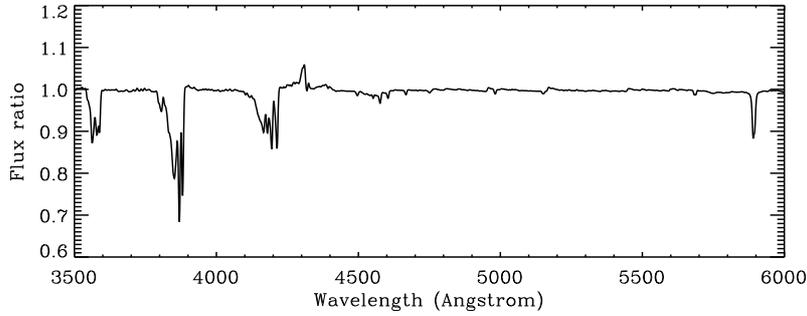}}
\caption{Spectral ratio between the modified CNONa and standard models. The strong features in the blue are related to CN bands.}
\end{figure}

\begin{figure}
\centerline{\includegraphics[bb=55 360 530 550,width=11cm]{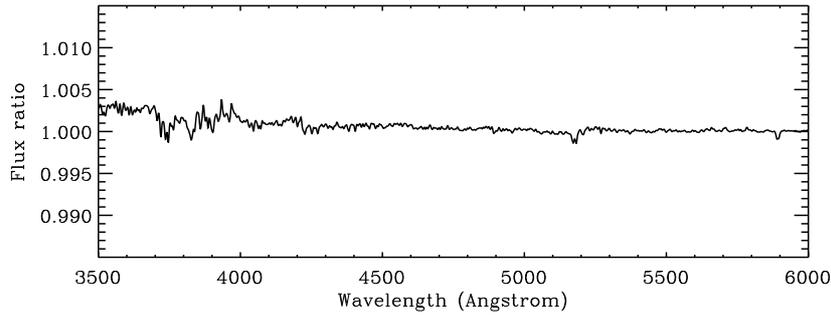}}
\caption{Spectral ratio between the He enhanced and modified CNONa models. The change of the continuum slope and the features related to H lines are a result of the hotter TO temperatures of the He enhanced model.}
\end{figure}


\section{Conclusions}

We explored the effects that primordial abundance changes of C, N, O and Na have on integrated spectra of a typical metal-rich galactic globular cluster. These variations affect appreciably only the Ca4227, G4300, ${\rm CN_1}$, ${\rm CN_2}$ and NaD metal indices. Ages, Fe-abundances and $Z$  
inferred from ${\rm H\beta}$-Fe5406 and ${\rm H\beta}$-[MgFe] diagrams (or the ${\rm H\delta_F}$ and,  
to a slightly smaller extent, the ${\rm H\gamma_F}$ counterpart) are confirmed once again to be robust. If the CNONa changes are accompanied by He enhancement, there might be a bias towards younger age determinations of up to 2\,Gyr. We refer the reader to \citet{coelho+11a} for the complete results.

The effect of these abundance changes on ages and metallicities derived via spectral fitting techniques remains to be
quantified. It might be particularly interesting to explore if these primordial abundance variations can explain the strong CN features observed in globular clusters in M31 (the author thanks Scott Trager for pointing this out).

\section*{Acknowledgements}

PC acknowledges the financial support by FAPESP via project 2008/58406-4 
and fellowship 2009/09465-0, and  
is particularly grateful to Fiorella Castelli, Piercarlo Bonifacio and the {\sc kurucz-discuss} mailing list
for the help with ATLAS and SYNTHE codes.


\label{lastpage}
\end{document}